# A novel method for unambiguous ion identification in mixed ion beams extracted from an EBIT

W. Meissl[1], M. C. Simon[1], J. R. Crespo López-Urrutia[2], H. Tawara[2],

J. Ullrich[2], HP. Winter[1], and F. Aumayr[1*]

[1] Institut für Allgemeine Physik, Technische Universität Wien, A-1040 Wien, Austria

[2] EBIT group, Max-Planck Institut für Kernphysik, D-69029 Heidelberg, Germany

## Abstract

A novel technique to identify small fluxes of mixed highly charged ion beams extracted from an Electron Beam Ion Trap (EBIT) is presented and practically demonstrated. The method exploits projectile charge state dependent potential emission of electrons as induced by ion impact on a metal surface to separate ions with identical or very similar mass-to-charge ratio.

* corresponding author

e-mail: aumayr@iap.tuwien.ac.at

Tel.: +43 1 58801 13430, Fax: +43 1 58801 13499

postal address: Institut f. Allgemeine Physik, TU Wien,

Wiedner Haupstr. 8-10/E134, A-1040 Vienna, Austria

# 1. INTRODUCTION

An Electron Beam Ion Trap (EBIT [1]) is a device to breed, store, investigate and also to extract highly charged ions (HCIs) [2-6]. In an EBIT, the ions of interest are produced by bombardment with a mono-energetic electron beam. Their motion is radially and axially confined by the electron beam space charge and potentials applied to a set of drift tube electrodes, respectively. During their confinement in the trap, the ions are brought consecutively to higher charge states by collisions with beam electrons (see e.g. [6-9] and refs. therein).

EBITs are used to investigate a wide variety of processes related to the physics of highly charged ions. Of fundamental importance are spectroscopic measurements of line radiation emitted from excited HCIs in order to test relativistic quantum mechanics, quantum electrodynamics (QED), electron correlation and nuclear physics (for recent work see e.g. [10-12]). More recently, ion beams extracted from an EBIT are utilized in a wide range of applications, from those associated with fusion plasma physics to those associated with nano-scale device fabrication [6-8])

Most EBITs utilize superconducting magnets which not only provide the necessary compression of the electron beam but also allow for cryopumping of the trap region. The base pressure in an EBIT is therefore quite low ($\leq 10^{-10}$ mbar). However, cryopumping also conserves a memory on previously used other working gases. Extracted ion beams therefore usually do not only contain ions of the desired working gas in different charge states but also a mixture of different ion species with comparable mass-to-charge ratios. Mass scans of ion beams extracted from an EBIT usually performed with an analyzing sector magnet can thus be quite complex, with the ions of interest not identifiable in a straight-forward manner.

We have devised a new method for characterizing extracted highly charged ion beams by means of the potential emission of electrons as induced by ion impact on a metal surface. In the following this method will be explained and demonstrated in a practical case.

## 2. EXPERIMENTAL METHOD

The details of the Heidelberg EBIT can be found elsewhere [9]. In its continuous mode of operation, also referred to as leaky or dc mode, ions that have enough thermal energy escape the trap by overcoming the electrostatic barrier at the confining drift tube held at constant potential. Extracted ions are replaced continuously by injected gas atoms. The extracted ion beam contains different ion species with various charge states. The ion beam is focused by an Einzel-lens and transported to the experimental chamber via a 90° analyzing magnet which is used to separate ions with desired mass-to-charge.

For the present study the central drift tube of the EBIT is biased to +6.4 kV, with the trap potential set to +210 V. 360mA emission current from the electron gun leads to a space charge potential of -250 V. That determines the ion extraction energy of 6.35 kV. The electron gun itself was biased to -7.5 kV, which together with a cathode bias of another -1.5 kV yields an electron beam energy of 15.1 kV, sufficient to produce bare $Ar^{18+}$, He-like $Xe^{52+}$ or Ne-like $W^{64+}$.

A typical mass over charge scan showing the number of transported ions versus the field of the analyzing magnet (already converted into a mass-to-charge ratio) for continuously extracted ions is shown in fig. 1. For this scan $^{129}Xe$ has been used as working gas, but the ion beam is apparently strongly contaminated and even dominated by residual gas ions (mainly oxygen and carbon) as well as by still present ions from $^{40}Ar$ (used as working gas shortly before). There is no possibility to distinguish between ions with identical mass over charge ratio with a scan like this. In our simple EBIT diagnostic device the ion beam enters a differentially pumped UHV chamber before hitting a clean metal surface under close to 30° angle of incidence (fig. 2). In our case we use a sputter-cleaned, single crystalline Au(111) target, but any other clean (polycrystalline) metal surface would do as well.

To detect the number of emitted electrons we apply a slightly modified version of the electron detection scheme described by Lemell et al. [13]. Electrons emitted from the ion-surface interaction region are extracted by a weak electric field through a highly transparent grid and accelerated onto a surface barrier type detector (Canberra PD 100-12-300 AM) biased at +25 kV. Ray-trace calculations performed for this geometry with the program SIMION showed that a field of about 100 V/cm is sufficient to collect all electrons with energies below 50 eV emitted into the half solid angle. Electron emission induced by a single projectile ion will be finished in a time much shorter than the time resolution of the applied detector electronics. Thus, $n$ electrons emitted due to a particular ion impact will be registered like *one* electron of $n \cdot 25$ keV rather than $n$ individual 25 keV electrons. The number of electrons emitted in a particular ion-impact event can therefore be deduced from the detectors pulse height distribution. More details on this electron number detection method and its appropriate evaluation can be found elsewhere ([13, 14] and refs. therein).

In general, if slow (impact velocity << 1a.u., corresponding to 25 keV/amu) HCIs collide with a metal surface, the majority of electrons is emitted by a process called potential emission (PE), where the potential energy of the ion (i.e. the sum of all ionization potentials) rather then it's kinetic energy (giving rise to the so-called kinetic emission – KE) dominates the total electron emission process [15, 16]. PE from metal surfaces has been found to strongly increase with the ion potential energy and hence it's charge state [5, 16].

It is the information on the precise number of the emitted electrons which allows us to separate ions with identical or very similar mass-to-charge ratio (see chap. 3). To this purpose the analyzing magnet is continuously scanned while for each individual ion impact event the number of emitted electrons is recorded in coincidence with the actual value of the magnetic field (fig. 3). For this technique typical projectile fluxes at the target must not exceed a few ten-thousand ions per second in order to avoid pulse pile-up, but may be as low as a few ions per second, which in fact makes the present method ideally suited for use with an EBIT ion source.

## 3. RESULTS

In fig. 4 the number of emitted electrons has been added as a second dimension to the mass over q scan shown in fig. 1. Each ion impact event is now characterized not only by the respective nominal charge to mass ratio (x-axis) but also by the amount of emitted electrons (y-axis). A series of distinct peaks are visible in this 2D-spectrum. While residual gas ions ($O^{q+}$, $C^{q+}$, etc.) are most prominent in intensity, their charge state (and potential energy) is comparably low, resulting in an electron emission yield of typically less than 10 $e^-$/ion. They can thus easily be discriminated from higher charged ions like the full series of $Xe^{q+}$ ions (q = 13 - 51). The series of HCIs from $^{40}$Ar is still visible and clearly separated from the $Xe^{q+}$ HCIs in the 2D-scan of fig. 4.

While the most prominent ions species could have been identified in a careful magnetic mass scan alone, the 2D scan in addition allows to identify trace ion species in the spectrum as well. Ions ejecting more than 150 electrons per impact event carry a higher charge than all Xe projectiles used. These ions were identified as $W^{q+}$ ($60 \leq q \leq 64$) HCIs from the cathode material of the EBITs electron gun, successively trapped and ionized in the EBIT itself. As three isotopes of tungsten are present (m=182, 184 and 186), the individual charge states cannot be separated anymore. Small peaks between the $Xe^{q+}$ peaks in the region $19 \leq q \leq 21$ point to ions with similar mass but slightly higher charge state than Xe (because of the slightly higher electron emission yield). They are identified as $Ba^{q+}$ ($20 \leq q \leq 25$) ions (also orginating from the e-gun). $Cu^{q+}$ ($15 \leq q \leq 23$) ions (material of the EBIT drift tubes) and traces of sulfur and fluorine HCIs are visible in the vicinity of the $Ar^{q+}$ peaks. Note that the $Ba^{q+}$ series has a different spacing on the mass-over-charge axis between its individual peaks than the $Xe^{q+}$ series. Due to this fact it is possible to distinguish even between two species, which have similar mass and charge, which would be impossible using a 1D electron number spectrum alone.

A "partial" $Xe^{q+}$ mass spectrum can be obtained (fig. 5) by properly selecting the region of interest as shown in the insert in fig. 5. Such a pure Xe spectrum is very convenient for an easy calibration of the relation between the field of the analyzing magnet and the mass-to-charge scale. The image also shows a distinct drop in rates for charge states higher than $Xe^{44+}$.

Apart from the choice of extraction parameters, this is also explained by the high ionization energies of the L shell and the fact that the trapping force in an EBIT increases linearly with q. Hence, the higher charge states are always underrepresented in an extraction spectrum from an EBIT in leaky mode, As the trap preferentially expels ions with lower charge states in a process called evaporative cooling. This also explains the presence of a population of tungsten in charge states from 60 to 64 with ionization potentials from 6.1 keV to 7.1 keV, and the lack of its lower charge states. Tungsten is accumulated in the trap very slowly, in a time in the order of minutes and therefore, all tungsten ions reach a very similar charge state.

In our case the target is mounted on a manipulator, so the HCIs of choice selected in this way can subsequently be used for performing experiments further downstream by simply retracting the target (fig. 1). In the case of a true mass-to-charge coincidence between two different ion species, the ratio between the two HCI-species can at least be quantified [17] (due to an almost 100% efficiency of HCI detection because of the high electron yields involved) and also optimized by tuning the ion source parameters.

Our technique not only provides a complete quantitative analysis of all ion beam fractions, it is also easily implemented, well suited to be incorporated into an existing beam-line and can be performed in a time comparable to that of a usual ion beam mass scan (typically 20 minutes).

**ACKNOWLEDGEMENTS**

This work has been supported by Austrian Science Foundation FWF (Project No. 17449) and was carried out within Association EURATOM-ÖAW. The experiments were performed at the distributed LEIF-Infrastructure at MPI Heidelberg Germany, supported by Transnational Access granted by the European Project HPRI-CT-2005-026015.

**FIGURE CAPTIONS**

Fig 1. (Color online) Typical mass scan of ions continuously extracted from an EBIT showing the number of transported ions versus the field of the analyzing magnet (converted into a mass-to-charge ratio). Unambiguously identified $Xe^{q+}$ peaks are shaded.

Fig 2. (Color online) Setup to identify slow highly charged ions by measuring the number of emitted electrons after ion impact on a metal surface

Fig 3. (Color online) Setup of the electronics (schematic).

Fig 4. (Color also in print version) For all ion impact events the number of emitted electrons is recorded together with the actual value of magnetic field of the analyzing magnet converted into ion charge-to-mass ratio. Total scanning time for this spectrum was 30 minutes.

Fig. 5. (Color online) A "partial" $Xe^{q+}$ mass spectrum as obtained from fig.3 by properly selecting the region of interest as shown in the insert.

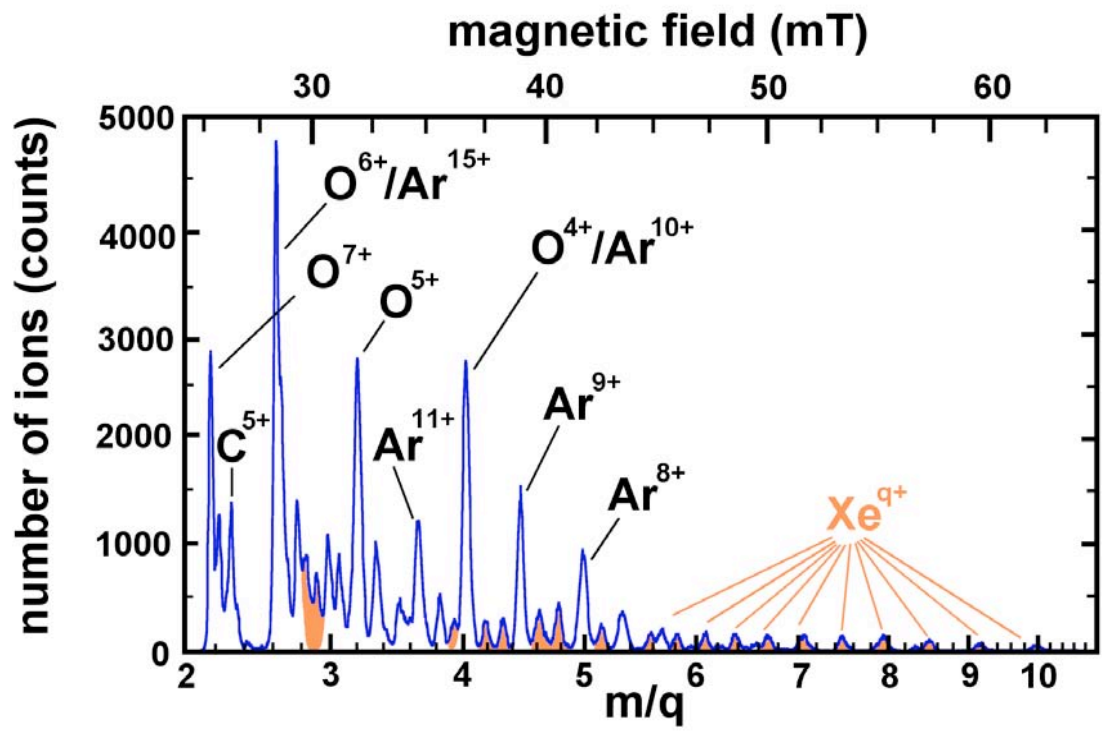

Fig. 1

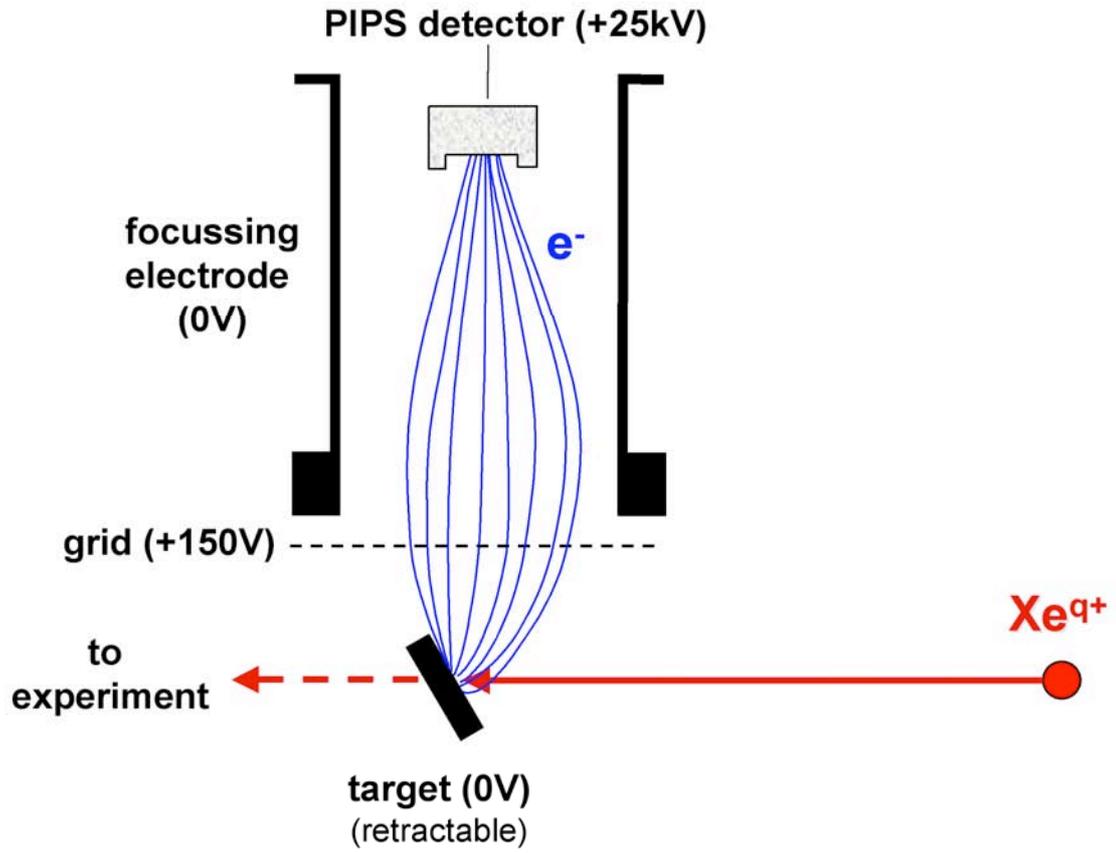

Fig. 2

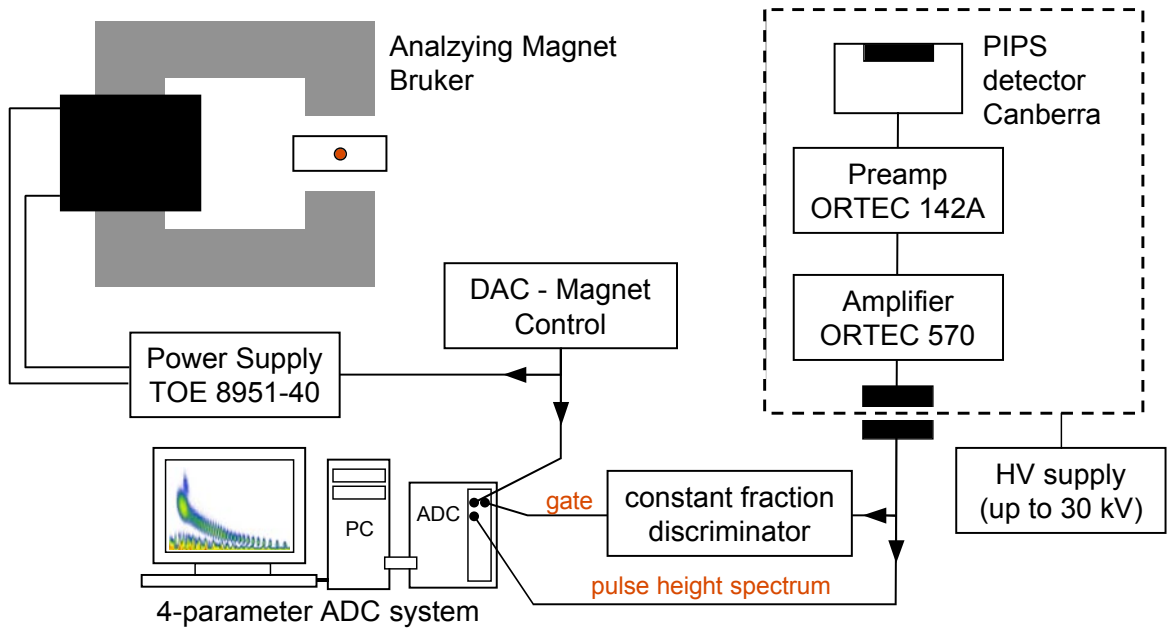

Fig. 3

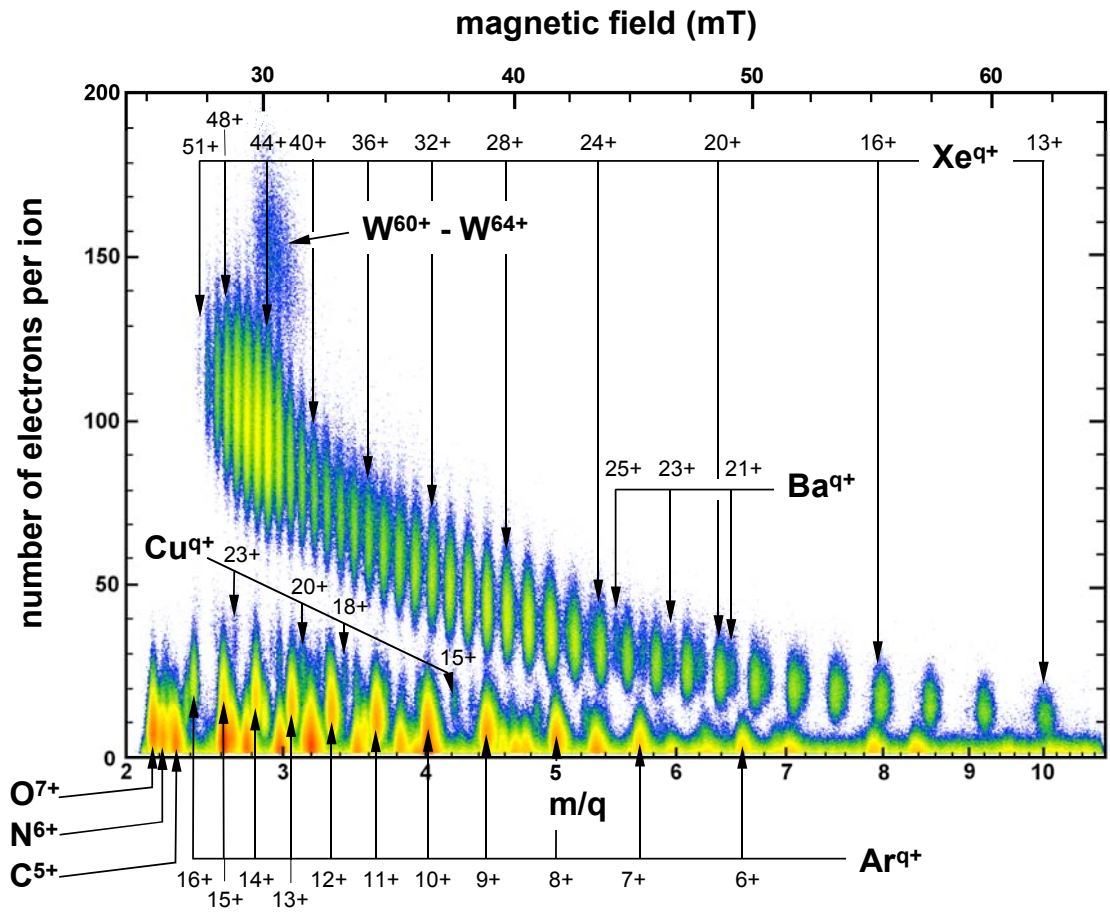

Fig. 4 (in color, width: 2 columns)

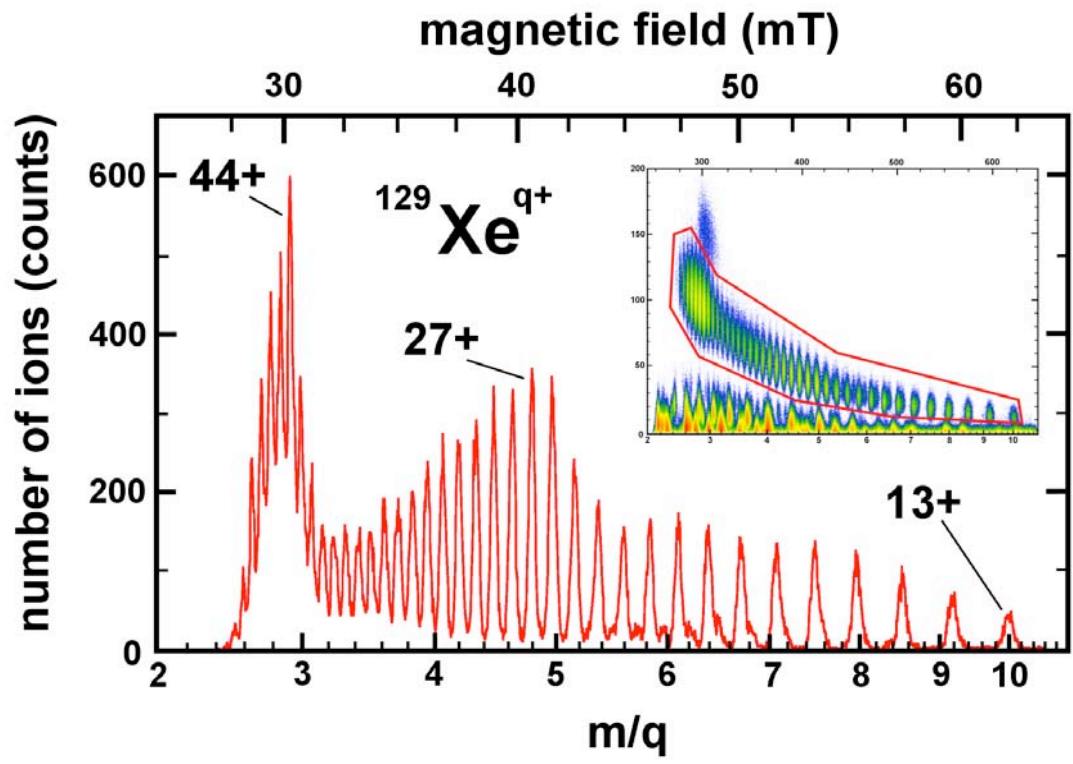

Fig. 5